\documentclass[twocolumn,aps,prl,showpacs]{revtex4}
\usepackage{epsfig}
\usepackage{graphicx}
\usepackage{amsmath}
\usepackage{amsfonts}
\usepackage{amssymb}
\usepackage{color}
\renewcommand{\Vec}[1]{{\bf #1}}
\definecolor{lred} {rgb}{.2,.5,1}

\begin{document} 
\title{Spin Hall effect in iron-based superconducting materials:\\
An effect of Dirac point}

\author{Sudhakar Pandey$^1$}
\author{Hiroshi Kontani$^1$}
\author{Dai S. Hirashima$^1$}
\author{Ryotaro Arita$^{2,3}$}
\author{Hideo Aoki$^{4,5}$}

\affiliation{$^1$Department of Physics, Nagoya University, Furo-cho,
 Nagoya 464-8602, Japan \\ 
$^2$Department of Applied Physics, University of Tokyo, Hongo, 
Tokyo 113-8656, Japan \\
$^3$PRESTO, Japan Science and Technology Agency (JST), Kawaguchi,
 Saitama 332-0012, Japan \\ 
$^4$Department of Physics, University of Tokyo, Hongo, Tokyo 113-0033, Japan\\
$^5$TRIP(JST), 5 Sanbancho Chiyoda-ku, Tokyo 102-0075, Japan}  


\begin{abstract}
We have theoretically explored the intrinsic 
spin Hall effect (SHE) in the iron-based superconductor family with a variety of materials.  
The study is motivated by an observation that, in addition to 
an appreciable spin-orbit coupling in the Fe $3d$ states, 
a character of the band structure in which 
Dirac cones appear below the Fermi energy  
may play a crucial role in 
producing a large SHE. 
Our investigation does indeed predict a substantially large 
spin Hall conductivity 
in the heavily hole-doped regime such as KFe$_2$As$_2$. 
The magnitude of the SHE has turned out to be 
comparable with that for Pt despite a relatively small spin-orbit coupling, which we identify to come from 
a  huge contribution from the gap opening induced by 
 the spin-orbit coupling at the Dirac point, which can become 
close to the Fermi energy for the heavy hole doping.    
\end{abstract}

\maketitle

There is a mounting interest in the 
spin Hall effect,  
where a transverse spin current 
(as opposed to charge current) 
 is induced in an electric field\cite{Vignale-Review-10}, which 
provides a most promising 
 way for manipulating spin degrees of freedom for 
 spintronic devices based on nonmagnetic materials. 
After its theoretical prediction  
in 1999 \cite{Hirsch-PRL-1999}, which was originally proposed  
in 1971 \cite{Dyakonov-Phys-Lett-1971}, 
investigations were initially focused 
mostly in semiconductors\cite{Kato-Science-2004}.
More recently, attention has been extended toward metallic systems 
\cite{Kimura-PRL-07-I,Kontani-JPSJ-07,Nagaosa-PRL-08,Kontani-PRL-08,Tanaka-PRB-08,Kontani-PRL-09,Tanaka-PRB-10,Fert-PRL-11,Morota-PRB-11,Tanaka-NJP-09,Guo-PRL-09},
following experimental reports of a huge spin Hall conductivity 
(SHC) $\sim 240 \hbar e^{-1} \Omega^{-1} {\rm cm}^{-1}$ 
in Pt \cite{Kimura-PRL-07-I}  
that is about $10^4$ times larger than the reported values in 
semiconductors \cite{Kato-Science-2004}.  
A key question then 
is how we can identify the materials that can exhibit large SHE.

Although SHE is driven by the spin-orbit interaction (SOI),  
understanding its mechanism
in {\it metallic} systems, especially intrinsic  vs extrinsic 
mechanisms, remains an open issue \cite{Vignale-Review-10}.
The intrinsic mechanism depends on the details of electronic
structure and predicted to be realized when the residual resistivity is mainly contributed by the randomness of the 
crystal \cite{Kontani-JPSJ-07,Nagaosa-PRL-08,Kontani-PRL-08,Tanaka-PRB-08,Kontani-PRL-09,Tanaka-PRB-10,Fert-PRL-11}. 
The extrinsic mechanism, on the other hand, is governed by external impurities in the host 
metallic system and becomes important  when the impurity atoms with $d-$ or $f-$ orbital degree of freedom give 
the dominant scattering\cite{Tanaka-NJP-09,Guo-PRL-09}.
Here we find that the intrinsic mechanism is dominant for the systems of our interest. 
This mechanism, apart from  accounting for the observed huge SHC in
Pt \cite{Kontani-JPSJ-07,Nagaosa-PRL-08,Tanaka-PRB-08,Kontani-PRL-09}, also predicts a 
variety of  materials, such as several $4d$ and $5d$ transition metals
\cite{Tanaka-PRB-08,Kontani-PRL-09}, as good candidates
for SHE. Interestingly, many of these predictions have been realized recently in experiments \cite{Morota-PRB-11}.

In this Rapid Communication we explore the possibility of SHE in 
the recently discovered iron-based superconducting materials, 
such as iron pnictides and chalcogenides.  
While the materials are now believed to be another class of 
high-$T_c$ superconductors after cuprates \cite{Wang-Science-11}, 
the reason why we look  
from the viewpoint 
of SHE is the following. These materials are   
typically {\it multiband} systems, where various $d$ orbitals are 
involved in the conduction bands \cite{Kuroki-2008}. 
The bands are entangled, 
namely, cross with each other with different orbital characters 
as a consequence of crystal symmetries in the iron-based materials, a situation that leads
to the occurrence of  
``Dirac cones" \cite{Anderson-Annalen-2011}.
We have also  
an appreciable spin-orbit coupling 
(SOC) for the  Fe $3d$ states and a quasi-two-dimensional nature of dispersion.  
Such features are expected to produce large SHE.   
For example, a large and quantized SHC is predicted 
in graphene when the Dirac cone becomes massive in presence of SOI\cite{Kane-PRL-05}.
Interestingly, the relatively strong strength of the SOC for the Fe $3d$ states ($72 meV$) \cite{Naito-PRB-10}
in comparison to the case of graphene($\simeq 4 meV$)\cite{Yao-PRB-07} is also favorable for experimental
realization of SHE in these materials.

This is our reasoning, and, if the Dirac points in these materials lead to a 
large SHE, then, apart 
from being a good candidate for spintronics, the materials may 
also provide an avenue 
for exploring the Dirac physics in the context of SHE in metallic systems 
that has recently drawn a surge of interests in case of 
topological insulators \cite{Hasan-RMP-10}.  
The present study does indeed predict a substantially 
large SHC for heavily hole-doped system such 
as KFe$_2$As$_2$, 
whose origin, as expected, lies in the SOC-induced gap at Dirac points that lie almost at
the Fermi level. We stress the Dirac-point-originated large SHE in the iron-based 
materials is distinct from other systems 
such as Pt and $4d$ and $5d$ 
transition metals, where 
substantially stronger SOC's govern the behavior of SHE.

\begin{figure}
\begin{center}
\vspace*{-2mm}
\hspace*{-5mm}
\psfig{figure=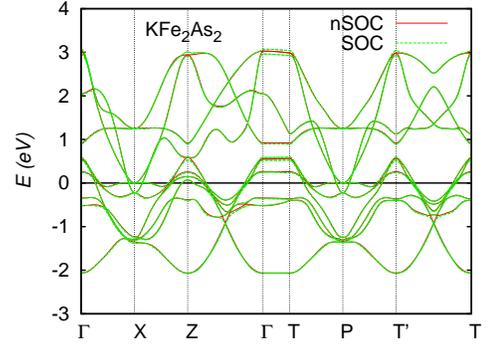,width=70mm}
\end{center}
\vspace*{-5 mm}
\caption{
Band structures with (SOC) and without (nSOC) spin-orbit coupling 
are shown for KFe$_2$As$_2$.  
The symmetry points are: $\Gamma=(0,0,0)$,
X$=(\pi,0,0)$, Z$=(\pi,\pi,0)$, T$=(0,0,\pi)$, P$=(\pi,0,\pi)$, and
T'$=(\pi,\pi,\pi)$.
In our convention, 
$E=0$ represents the chemical potential 
$\mu$, and $x-$ and $y-$ are directed along the nearest neighbor Fe atoms.}  
\label{fig:band-DOS}
\vspace{-3mm}
\end{figure}

We consider the realistic band structure of various types of the iron 
compounds with two Fe atoms 
per unit cell by incorporating the SOC within an effective tight-binding (TB)
Hamiltonian.
The Wannier basis of the TB model has been constructed as follows.
We first performed a density-functional calculation, where we used the 
exchange correlation functional proposed by Perdew {\it et al.}
\cite{PBE-PRL-1996}, and the augmented plane wave and local orbital (APW+lo) method including the SOC 
as implemented in the
WIEN2k code\cite{WIEN2k-Web,SO-PRB-01}.  We then constructed the TB model
using the WIEN2Wannier\cite{Kunes-Comp-Phys-Comm-10} and the
wannier90 \cite{wannier90-Comp-Phys-Comm-08}. In order to preserve the local
symmetry of the Wannier functions, we perform the
so-called one-shot Wannierization \cite{MaxLoc-Vanderbilt-PRB-1997-01}.
Our approach provides a good description of several common electronic
features in these  materials  \cite{Arita-JPSJ-10} such as  
an effective $d$-electron bandwidth of $4.5 - 5.0$ eV, 
and the dominant orbital 
character of electronic states etc., 
which are almost independent of the SOC, as can 
be seen in Fig. \ref{fig:band-DOS}, and  
expected from the weak SOC for the Fe $3d$ electrons.
However, we shall discuss later that some
SOC-induced band features in fact play a crucial role in the SHE.

Intrinsic mechanism based SHE is investigated with the linear-response theory\cite{Streda-JPC-1982}
in the presence of local impurities that give rise to a finite residual resistivity. 
At $T=0$, SHC consists of two parts, 
$\sigma_{xy}^z=\sigma_{xy}^{z{\rm I}}+\sigma_{xy}^{z{\rm II}}$, where

\begin{equation}
\sigma_{xy}^{z{\rm I}} =\frac{1}{2 \pi N} 
\sum_{\Vec{k}} {\rm Tr} \left [\hat J_x^S \hat G^R \hat J_y^C \hat G^A \right ]_{\omega=0}
\end{equation}
represents the ``Fermi-surface term", while
\begin{eqnarray}
\sigma_{xy}^{z{\rm II}} &=& 
 - \frac{1}{4 \pi N} \sum_{k} \int_{-\infty}^0 
d\omega  {\rm Tr} \left [\hat J_x^S \frac {\partial \hat G^R} {\partial \omega} \hat J_y^C \hat G^R \right.
\nonumber  \\
&-&
\left.
\hat J_x^S \hat G^R  \hat J_y^C  \frac {\partial \hat G^R} {\partial \omega} - 
\langle R \leftrightarrow A \rangle 
\right ]
\end{eqnarray}
represents the ``Fermi-sea term".  
Here the charge-current operator is given by 
 $\hat{J_{\zeta}^C} = -e \partial \hat H/\partial {\Vec{k}}_{\zeta}$ 
 with $- e   (< 0$) being the electronic charge, and 
$\zeta=x,y$, while the $\sigma_z$-spin 
 current operator is given by
  $\hat J_{\zeta}^S = (-1/e) \{ \hat J_{\zeta}^C, \hat \sigma_z \}/2 $.
The retarded (advanced) Green's function $\hat G^R (\hat G^A)$ are  
given by 
$\hat G^{R/A} (\Vec{k}, \omega)  = 1/(\omega + \mu -\hat{H} \pm i \hat{\Gamma})\;$. 
 Here $\hat{H}$ is the effective TB Hamiltonian, a $20 \times 20$ matrix
 spanned by the two Fe atoms per unit cell in the presence of SOC,
 and $\mu$ the chemical potential.  $\hat{\Gamma}$, the 
 damping due to local impurities, is treated 
with the T-matrix approximation, 
 $\hat{\Gamma} = ( n_{\rm imp}/{2 i}) 
[ \hat{T} (-i 0) - \hat {T} (+i 0) ]$, where $n_{\rm imp}$ 
 is the impurity concentration, and 
 $\hat {T} (\pm i 0)= \hat {I}/ [1-\hat{I} \hat{g}(\pm i 0)]$, with 
 $\hat{g}(\omega) = \sum_{\Vec{k}} G (\Vec{k}, \omega)$ representing the 
T-matrix for a single impurity with an  impurity potential 
 ${\hat I}$.  For simplicity we consider a constant and orbital-diagonal
 impurity potential of strength $I$, i.e.,
 ${\hat I}_{\alpha,\beta}=I\delta_{\alpha,\beta}$.

 In the quantitative investigations that follow, our results for SHC 
 are expressed in unit of $|e|/2\pi a$,  
 where $a$ is the interlayer spacing. For $a=6 \AA$,  
 $|e|/2\pi a \sim 645  \hbar e^{-1} \Omega^{-1} {\rm cm}^{-1}$.
 Unless otherwise mentioned, we consider $I=6$ eV, and $n_{\rm imp}=0.01$.

\begin{figure}
\begin{center}
\vspace*{-50mm}
\hspace*{-40mm}
\psfig{figure=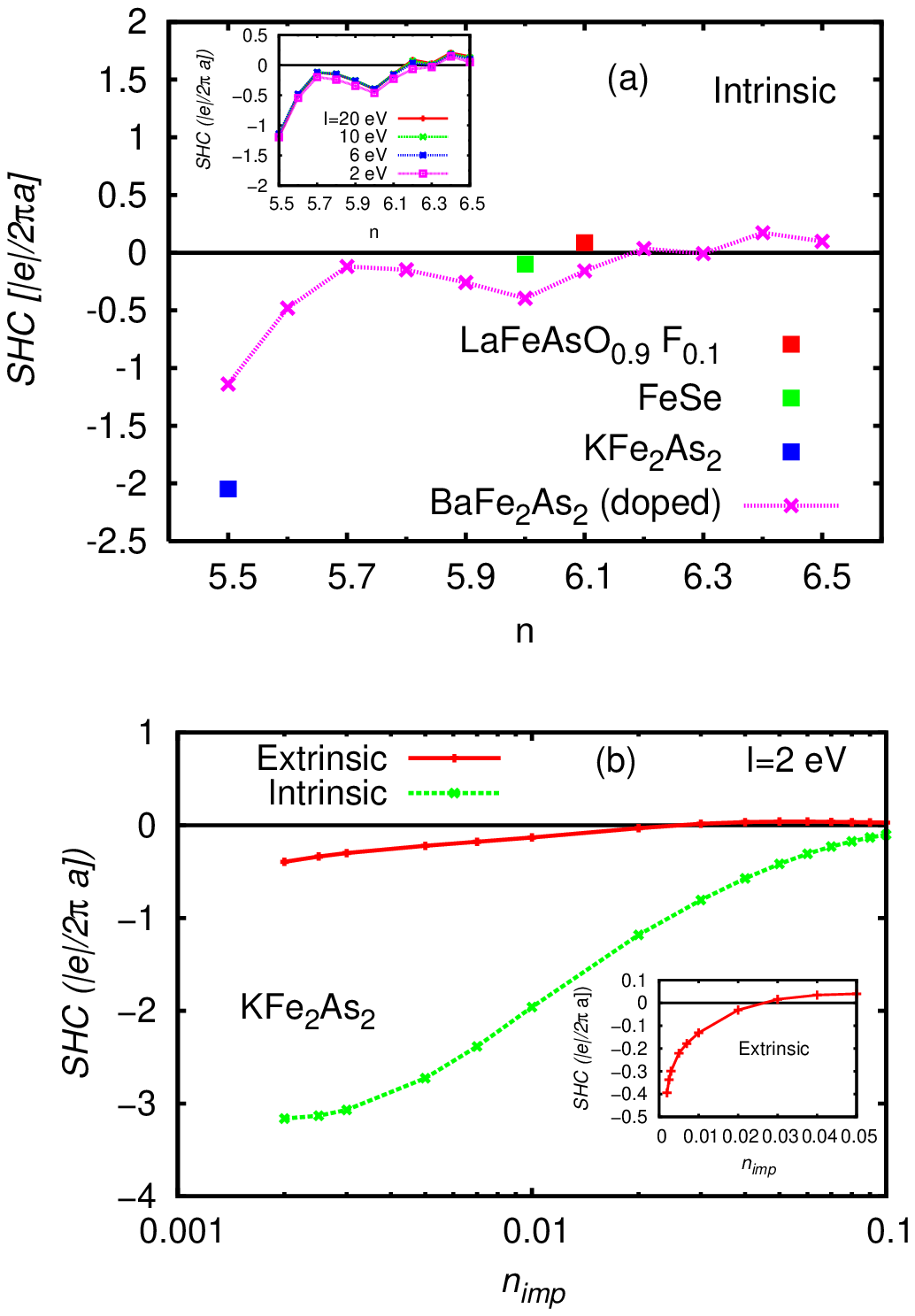,width=130mm}
\vspace*{-56mm}
\end{center}
\caption{(a) Intrinsic mechanism based results for SHC  
for various systems (symbols).  
The curve is the doping dependence 
in the rigid-band approximation 
for the band structure of BaFe$_2$As$_2$. Inset shows the 
dependence of the result on the strength of impurity potential. 
(b) Relative contributions of the intrinsic ($\sigma_{xy}^{\rm I}$) and extrinsic ($\sigma_{xy}^{ss}$) mechanism to SHC as a function of impurity concentration ($n_{imp}$), as shown 
for KFe$_2$As$_2$. Here inset demonstrates the rapid increase 
of the extrinsic mechanism in the limiting case  $n_{imp} \rightarrow$ 0} 
\label{fig:SHC-All}
\vspace*{-5mm}
\end{figure}

Figure \ref{fig:SHC-All} (a) shows the results for SHC 
for various materials, KFe$_2$As$_2$, FeSe, and LaFeAsO$_{0.9}$F$_{0.1}$, 
obtained with the {\it ab initio} band structures. 
The dominant contribution is found to arise from the 
Fermi-surface part, i.e., $\sigma_{xy}^z \simeq \sigma_{xy}^{z{\rm I}}$, 
for all the materials considered. While 
a large SHC arises in KFe$_2$As$_2$, 
a heavily hole-doped system,  
it becomes vanishingly small in 
undoped FeSe and weakly electron-doped 
LaFeAsO$_{0.9}$F$_{0.1}$.  
Interestingly, the large 
magnitudes of SHC $\sim 1300 \hbar e^{-1} \Omega^{-1} {\rm cm}^{-1}$ 
in KFe$_2$As$_2$ with a residual resistivity 
$\rho \sim 10 \mu \Omega \rm cm$
is comparable to that 
($\sim 1000 \hbar e^{-1} \Omega^{-1} {\rm cm}^{-1}$)
predicted for Pt with a substantially larger SOC in the same metallic 
regime, where the observed SHC
 $\sim 240 \hbar e^{-1} \Omega^{-1} {\rm cm}^{-1}$
 is reproduced at higher 
$\rho \sim 100 \mu \Omega \rm cm$ 
\cite{Kontani-JPSJ-07,Tanaka-PRB-08,Kontani-PRL-09}.
We  
trace the origin of 
the large SHC in KFe$_2$As$_2$ in a special band feature 
later.
  
\begin{figure}
\begin{center}
\vspace*{-27mm}
\hspace*{-30mm}
\psfig{figure=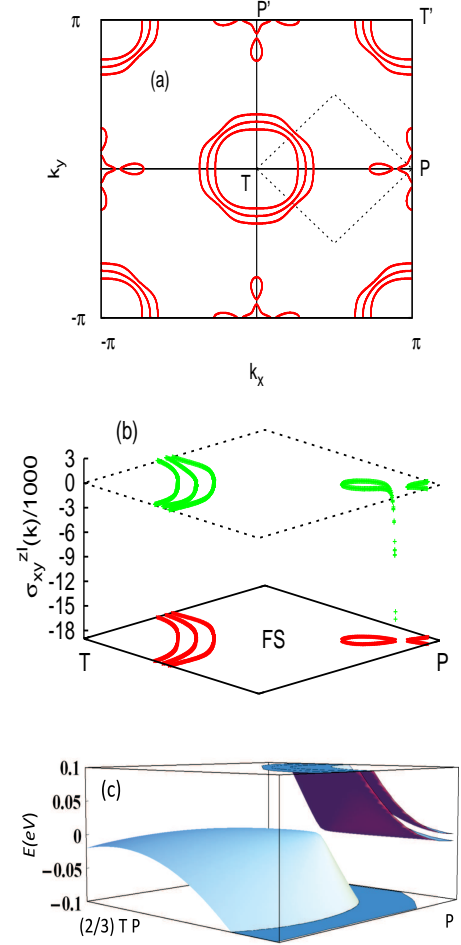,width=138mm}
\vspace*{-52mm}
\end{center}
\caption{(a) Fermi surface for KFe$_2$As$_2$ shown on the $k_z=\pi$ plane.  
(b) ${\Vec{k}}$-dependent contribution from the states on 
the Fermi surface (red lines) 
to SHC (green),
displayed for a part of Fermi surface  (dotted region in (a)). 
Here dotted region represents the $\sigma_{xy}^{z{\rm I}}(\Vec{k})=0$ plane.
(c) Gapped Dirac cone dispersion near P. 
Electronic states make the huge contribution to SHC (b) in the close vicinity of the gap.}
\label{fig:band-FS-SHC-Surf-122}
\vspace*{-5mm}
\end{figure}

Since the SHE is found to be significantly affected by the doping level,  
we next investigate the effect of carrier doping by  
varying the band filling over a range 
($n=5.5 - 6.5$) that goes from electron doped ($n>6$) to 
hole doped ($n<6$) sides as relevant to the family of 
these materials. We follow  
a rigid-band approach  
using the
band structure of BaFe$_2$As$_2$ that is known to 
cover both electron doping as in 
Ba(Fe$_{1-x}$Co$_x)_2$As$_2$ and 
 hole doping as in Ba$_{1-x}$K$_x$Fe$_2$As$_2$.  
 Here we only consider the Fermi-surface part, which is found to be dominant as discussed  above.     
The result, a curve in Fig. \ref{fig:SHC-All} (a), 
exhibits a large SHC in 
the heavily hole-doped regime ($n \simeq 5.5$), while 
the magnitude is small in the weakly hole-doped and 
electron-doped regimes.  
The overall filling dependence agrees with 
the results for KFe$_2$As$_2$ ($n=5.5$), FeSe ($n=6.0$), 
and LaFeO$_{0.9}$F$_{0.1}$ ($n=6.1$), and 
highlights the presence of a common feature. 
Small deviations from the rigid-band result 
should be due to deformations in the electronic structures, such as the 
stronger three dimensionality in BaFe$_2$As$_2$.  

As for the dependence on the  impurity potential strength, 
we have checked, as shown in the inset of Fig. \ref{fig:SHC-All} (a), 
that, when we vary the strength over 
$(I\sim 2-20)$ eV, the SHC, including its filling dependence, 
is rather insensitive to the strength of impurity 
potential\cite{NakamuraArita}.  This hallmarks an intrinsic nature of SHE in these materials.

In order to further support our prediction for the intrinsic origin of SHE, 
we now estimate the contribution of extrinsic mechanism by focusing on
the {\it skew} scattering term\cite{Tanaka-NJP-09}, $\sigma_{xy}^{ss} = \frac{n_{imp}}{2} \frac{1}{2\pi} \sum_{l,m,n,o} [ B_{m,n} T^R_{n,l} A_{l,o} 
T^A_{o,m} + c.c ]$ with $A_{l,m}= \sum_k [G^R J_y^C G^A]_{l,m}$ and $B_{l,m}= \sum_k [G^A J_x^S G^R]_{l,m}$,  
which is predicted to be dominant in the limit of dilute impurities\cite{Tanaka-NJP-09}.  
As shown in Fig. \ref{fig:SHC-All} (b),   
the contribution of
skew scattering is nearly smaller by an order of magnitude in comparison to that of the intrinsic part for $n_{imp}\sim 0.01$. For $n_{imp} < 0.01$, while
intrinsic contribution tends to saturate, the skew scattering contribution increases rapidly in proportion to $1/n_{imp}$.
The two contributions become comparable in the regime $n_{imp} < 0.001$, where  
$\rho \le 1\mu\Omega cm$. 
Since such a high quality sample is very difficult to prepare experimentally, we believe that the intrinsic contribution is dominant in "realistic" clean samples with $\rho \ge 10\mu\Omega cm$.

In the following our discussion will be based on the intrinsic mechanism with a focus on the heavily hole-doped system KFe$_2$As$_2$ 
that exhibits the large SHC. 
In view of the dominant contribution from the Fermi surface part ($\sigma_{xy}^{z{\rm I}}$) we explore if certain special band features around 
the Fermi energy are responsible.  
We consider the
$\Vec{k}$-dependent contribution of electronic states on
the Fermi surface to SHC,  
 $\sigma_{xy}^{z{\rm I}}(\Vec{k}) = 
\frac{1}{8} \sum_{k'= \pm k_x, \pm k_y, \pm k_z}
 {\rm Tr} [(\hat J_x^S \hat G^R \hat J_y^C \hat G^A - \hat J_y^S \hat G^R 
 \hat J_x^C \hat G^A)/2]_{k',\omega=0}$ where we are averaging 
over $xy$ and $yx$ components etc., with
 $\frac{1}{2\pi N} \sum_\Vec{k} \sigma_{xy}^{z{\rm I}} (\Vec{k})$
 providing the net contribution of 
 $\sigma_{xy}^{z{\rm I}}$\cite{Kontani-JPSJ-07}.

\begin{figure}
\begin{center}
\vspace*{-52mm}
\hspace*{-16mm}
\psfig{figure=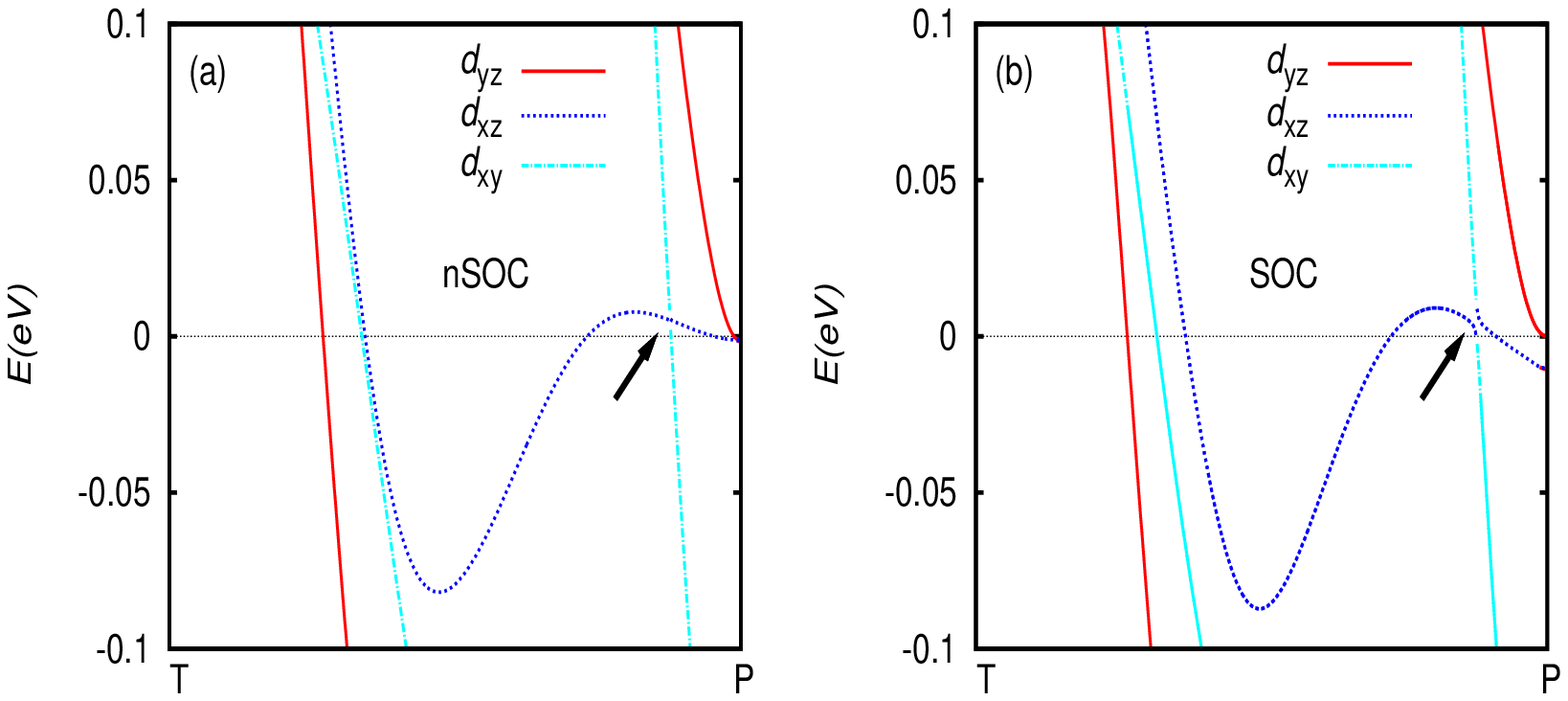,width=110mm}
\vspace*{-70mm}
\end{center}
\caption{Band dispersions for  KFe$_2$As$_2$ around the Fermi level ($E=0$)
without SOC (a) 
and with 
SOC (b).  
The Dirac point, 
around which dominant contributions to SHC arise 
as shown in Fig. \ref{fig:band-FS-SHC-Surf-122}, 
is indicated with arrows. The small gap ($\ll$ SOC) 
is reflected more clearly in Fig 3(c) due to tilted nature 
of the Dirac cone.}
\label{fig:band-T-P}
\vspace*{-5mm}
\end{figure}

We consider only a part of the Brillouin zone, as shown by the
the dotted area in Fig. \ref{fig:band-FS-SHC-Surf-122} (a), as 
the other parts can be deduced from the symmetry.  
For clarity we consider the $k_z=\pi$ plane, as 
other planes give qualitatively 
similar contributions due to quasi-2D nature of the Fermi surface  that 
is also apparent from the almost flat dispersion near the Fermi level
along $\Gamma-T$ direction (Fig. \ref{fig:band-DOS}). 
Figure \ref{fig:band-FS-SHC-Surf-122} (b) shows that a huge
contribution to SHC comes from a small region on the hole pocket  
near P,  
while the contributions from the rest part are not only relatively much smaller but also tend to cancel with each other. 
Therefore, origin of the large SHC is governed by the 
states on the hole pocket near P.

 We can actually 
identify, in Figures \ref{fig:band-FS-SHC-Surf-122} (c) and \ref{fig:band-T-P}, 
that these electronic
states lie in the close vicinity of a small SOC-induced gap in the 
Dirac cone.   
This is also the case with our analysis for the filling dependence in BaFe$_2$As$_2$ with a rigid band, 
where we find a similar origin behind the large SHC 
in the heavily hole-doped regime $(n \simeq 5.5)$, 
in which the Fermi level approaches the Dirac point.  Concomitantly, 
such Dirac-like points near the Fermi 
level are absent in FeSe and LaFeAsO$_{0.9}$F$_{0.1}$. 
Thus the large SHC when the Fermi level lies close to the 
SOC-induced gap at the 
Dirac point renders the iron-based systems a Dirac electron
system 
such as graphene\cite{Kane-PRL-05}.
 
Such Dirac-like points arise due to crossing of two bands  
with almost linear dispersions and 
different orbital characters 
as a consequence of crystal symmetries in the iron-based materials. 
In the 
present case,  while  one of the crossing bands has dominantly $xz$ orbital character,
the other band has $xy$ character (Fig. \ref{fig:band-T-P}). 
The band crossing is a consequence of the opposite parities of the two orbitals 
with respect to reflection $y \rightarrow -y$. In the presence of an SOI, the gap opens 
because  the 
two orbitals are then coupled by the $x-$component of the 
orbital angular momentum 
${\hat l_x}$; $\langle xz |{\hat l_x}| xy \rangle = i$ \cite{Tanaka-PRB-08}.

We also study the temperature dependence of SHC in KFe$_2$As$_2$. As expected from the small SOC-induced gap at the Dirac point, the SHC decreases monotonically,
and at room temperature ($300K$), it reduces to about $60\%$ of its $T=0$ value that is still significantly large and favorable for device application.

Although it is an important future problem to study the role of correlation effects on the
 SHC, we expect that it is less important in
Fe-based superconductors as follows:
The Coulomb interaction is much smaller than the
bandwidth according to the first principle study by Miyake {\it et al.}
\cite{Arita-JPSJ-10}, consistently with the small mass-enhancement factor $m^*/m=2 \sim 3$.
 Note that the SHC is independent of the mass-enhancement 
\cite{Tanaka-PRB-10}. 
Although spin fluctuations can sensitively influence transport
phenomena\cite{Kontani-Rep-Prog-Phys-08},
spin fluctuations in KFe$_2$As$_2$ are small according to $1/T1$
measurements, due to the bad nesting of the Fermi surface.
Thus we expect that the giant SHC in KFe$_2$As$_2$, which is
the most important result of our study, will be reliable.

In summary, we have theoretically explored the possibility of SHE  
in a variety of the  
iron-based superconducting materials.
We reveal that a substantially large SHE 
arises in heavily hole-doped
materials of 122 family 
such as KFe$_2$As$_2$, whose magnitude is even comparable with that in Pt. 
The large SHE is found to originate from 
a huge contribution from the electronic states in the vicinity of the SOC- 
induced gap at the Dirac points that  
lie close to 
the Fermi level in the heavily hole doped case. 
We hope that our study will stimulate an experimental investigation due to relative easiness in synthesis 
and higher sample quality of the 122 family of materials.

S. P. and R. A. are supported by the JSPS Postdoctoral Fellowship for Foreign 
Researchers,   
and the JST PRESTO program, respectively.

\end{document}